\documentclass[superscriptaddress,notitlepage,twocolumn,floatfix]{revtex4-1}

\usepackage{graphicx}
\usepackage{dcolumn}
\usepackage{bm}
\usepackage{amsmath}
\usepackage{amsfonts}
\usepackage{caption}
\usepackage{subcaption}
\usepackage[section]{placeins}
\usepackage{hyperref}

\begin{document}

\title{Fisher Information and Atomic Structure}

\author{K.Ch. Chatzisavvas}   \email{kchatz@auth.gr}
\affiliation{Department of Theoretical Physics, Aristotle University of Thessaloniki, 54124 Thessaloniki, Greece}
\affiliation{Department of Informatics and Telecommunications Engineering, University of Western Macedonia, 50100 Kozani, Greece}
\author{S.T. Tserkis}   \email{stserkis@physics.auth.gr} 
\affiliation{Department of Theoretical Physics, Aristotle University of Thessaloniki, 54124 Thessaloniki, Greece}
\author{C.P. Panos}    \email{chpanos@auth.gr}
\affiliation{Department of Theoretical Physics, Aristotle University of Thessaloniki, 54124 Thessaloniki, Greece}
\author{Ch.C. Moustakidis}  \email{moustaki@auth.gr}
\affiliation{Department of Theoretical Physics, Aristotle University of Thessaloniki, 54124 Thessaloniki, Greece}

\date{December 18, 2013}

\begin{abstract}

We present a comparative study of several information and statistical complexity measures in order to examine a possible correlation with certain experimental properties of atomic structure. Comparisons are also carried out quantitatively using Pearson correlation coefficient. In particular, we show that Fisher information in momentum space is very sensitive to shell effects, and is directly associated with some of the most characteristic atomic properties, such as atomic radius, ionization energy, electronegativity, and atomic dipole polarizability. Finally we present a relation that emerges between Fisher information and the second moment of the probability distribution in momentum space i.e. an energy functional of interest in (e,2e) experiments.

\end{abstract}

\maketitle


\section{Introduction}

A large amount of work has been published so far, investigating the atomic structure through information-theoretic tools and statistical complexity measures \cite{Massen, Chatzisavvas, Nikolaidis, Moustakidis,Tserkis, Lopez95, Sanudo, Montogomery, AnguloAntolin, AnguloSen, Patil, Sagar, Nagy, Szaboa, Antolin, Borgoo, Feldman, Yamano, Garbaczewski, Plastino}. These information and statistical complexity measures are, in essence, functionals of the studied density distributions, and therefore in our case, functionals of the atomic density distributions (probability distributions of electrons in atoms), in both momentum and position spaces. 

Preliminary studies were focused separately in position and momentum space densities\cite{Romera}, while later the interest shifted to ``total" or ``net" information measures, attempting to merge the results of both spaces (e.g. the sum of position and momentum spaces for the Shannon information entropy, and the product of the corresponding results for the Fisher information)\cite{Sen}.

In the present work we study separately various information and complexity measures in position and momentum spaces. Of course the ``total" measures, in the sense described above, satisfy some tempting and often desirable properties, e.g. they are dimensionless, but from a physical point of view each space (position and momentum) carries different information about the atomic structure. So, it is worth to examine which of the above measures resembles more sensitively and accurately the periodicity of the atomic structure i.e. carries more information. Thus, we present Shannon information entropy, Fisher information, and disequilibrium, together with LMC complexity and Fisher-Shannon plane, in both position and momentum spaces, as functions of the atomic number $Z$. We also compare our results with experimental data i.e. atomic radius, first ionization potential, electronegativity, and atomic dipole polarizability. An attempt is made to connect the Fisher information with (e,2e) experiments.

\section{Information Measures}\label{Information Measures}

The Shannon information entropy \cite{Shannon} in position space, $S_r$, is defined as

\begin{equation}
S_r=-\int \rho({\bf r})\ln{\rho({\bf r})}  \,d{\bf r},
\end{equation}

\noindent where $\rho({\bf r})$ is the electron density distribution normalized to unity. The corresponding information entropy in momentum space, $S_k$, is given by

\begin{equation}
S_k=-\int n({\bf k})\ln{n({\bf k})}  \,d{\bf k},
\end{equation}

\noindent where $n({\bf k})$ is the momentum density distribution normalized to unity. In position-space, $S_r$ determines the extent of electron delocalization, since a highly localized $\rho({\bf r})$ is associated with a diffuse $n({\bf k})$, leading to low $S_r$ and high $S_k$ and vice versa. The net Shannon information entropy takes the form

\begin{equation}
S_T=S_r+S_k.
\end{equation}

The Fisher information \cite{Fisher, Frieden} in position, $I_r$, is defined as

\begin{equation}
I_r=\int \frac{| \nabla \rho({\bf r})|^{2}}{\rho({\bf r})} {\rm d}{\bf r} \label{I-r-1},
\end{equation}

\noindent and the corresponding measure in momentum space is given by

\begin{equation}
I_k=\int \frac{| \nabla n({\bf k})|^{2}}{n({\bf k})}{\rm d}{\bf k}.
\label{I-k-1}
\end{equation}

The individual Fisher measures are bounded through the Cramer-Rao inequality according to $I_r \geq \frac{1}{V_r}$ and $I_k \geq \frac{1}{V_k}$ \cite{Rao, Stam}, where $V_r$ and $V_k$ are the corresponding spatial and momentum variances respectively. In position space, Fisher information measures the sharpness of probability density and for a Gaussian distribution is exactly equal to the variance \cite{Frieden}. A sharp and strongly localized probability density  gives rise to a larger value of Fisher information in position space. Moreover, the Fisher measure has the desirable properties, i.e. it is always positive and reflects the localization character of the probability distribution more sensitively than the Shannon entropy \cite{Garbaczewski, Carroll}. 

The net Fisher information is defined by

\begin{equation}
I_T=I_r I_k.
\end{equation}

Fisher information is also intimately related to the Shannon information entropy via de Bruijn identity \cite{Stam, Cover}

\begin{equation}
\frac{\partial}{\partial t}S(x+\sqrt{t}z)\bigg|_{t=0} =\frac{1}{2}I(x),
\end{equation}

\noindent where $x$ is a random variable with a finite variance with a density $f(x)$, and $z$ an independent normally distributed random variable with zero mean and unit variance.

The disequilibrium (also know as information ``energy" or Onicescu information) $D$ as defined in \cite{LMC, Onicescu}, for position and momentum spaces, is

\begin{align}
D_r &= \int \rho^2({\bf r}) {\rm d}{\bf r}, \\
D_k &= \int n^2({\bf k}) {\rm d}{\bf k},
\end{align}

\noindent while for the net disequilibrium $D_T$ we employ the definition

\begin{equation}
D_T=D_r D_k.
\end{equation}

The statistical measure of complexity \cite{LMC, Catalan} is defined by the formula

\begin{align}
C_r &= e^{S_r}\cdot D_r, \\
C_k &= e^{S_k}\cdot D_k.
\end{align}

\noindent while the net LMC complexity takes the form 

\begin{equation}
C_T=C_r C_k.
\end{equation}

Finally Fisher-Shannon plane \cite{RD} in position and momentum space in D-dimensions is defined as

\begin{align}
P_r &= \frac{1}{D} J_r I_r, \\
P_k &= \frac{1}{D} J_k I_k,
\end{align}

\noindent where $J_r=\frac{1}{2 \pi e} e^{\frac{2}{D}S_r}$ and $J_k=\frac{1}{2 \pi e} e^{\frac{2}{D}S_k}$, and the corresponding net measure is given by

\begin{equation}
P_T=P_r P_k.
\end{equation}

In the present work we consider accurate spin-independent atomic wave functions obtained by Bunge et al. \cite{Bunge}, by applying the Roothaan-Hartree-Fock method (RHF) to calculate analytical self-consistent-field (SCF) atomic wave functions. In this approach the radial atomic orbitals are expanded as a finite superposition of primitive radial functions

\begin{equation}
R_{nl}(r)=\sum_j C_{jnl} S_{jl}(r),
\end{equation}

\noindent where the normalized primitive basis $S_{jl}(r)$ is taken as a Slater-type orbital set

\begin{equation}
S_{jl}(r)=N_{jl} r^{n_{jl}-1}e^{-Z_{jl}r}.
\end{equation}

\noindent The normalization factor $N_{jl}$ is given by

\begin{equation}
N_{jl}=(2Z_{jl})^{(n_{jl}+1/2)}/[(2n_{jl})!]^{1/2},
\end{equation}

\noindent where $n_{jl}$ is the principal quantum number, $Z_{jl}$ is the orbital exponent, and $l$ is the azimuthal quantum number.

The atomic wave functions in the momentum space are related to the coordinate wave functions by

\begin{equation}
{\tilde \phi}_{nlm}({\bf k})=\frac{1}{(2\pi)^{3/2}} \int e^{-i{\bf kr}} \phi_{nlm}({\bf r}) d{\bf r},
\end{equation}

\noindent where $nlm$ denote the usual quantum numbers characterizing atomic states. The radial momentum-wave function defined as

\begin{equation}
{\tilde \phi}_{nlm}(k)=(-i)^{l}{\tilde R}_{nl}(k)Y_{lm}(\Omega_k),
\end{equation}

\noindent is related to the radial wave function in coordinate space through

\begin{equation}
{\tilde R}_{nl}(k)=\sqrt{\frac{2}{\pi}} \int_{0}^{\infty} r^2 R_{nl}(r)j_{l}(kr) dr,
\end{equation}

\noindent where $j_l(kr)$ is a spherical Bessel function. The radial wave functions in momentum space ${\tilde R}_{nl}(k)$ are written as

\begin{equation}
{\tilde R}_{nl}(k)=\sum_j C_{jnl} {\tilde S}_{jl}(k),
\end{equation}

\noindent in term of the RHF functions ${\tilde S}_{jl}(k)$ in momentum space, related to $S_{jl}(r)$ through

\begin{equation}
{\tilde S}_{jl}(k)=\sqrt{\frac{2}{\pi}}  \int_{0}^{\infty} r^2 S_{jl}(r)j_{l}(kr) dr.
\end{equation}

In the present work we consider $\rho({\bf r})$ and $n({\bf k})$ to be spherically symmetric functions.

\section{Results and Discussion}\label{Results and Discussion}

\begin{figure}[!htb]
\centering
  \includegraphics[height=20cm,width=8.5cm]{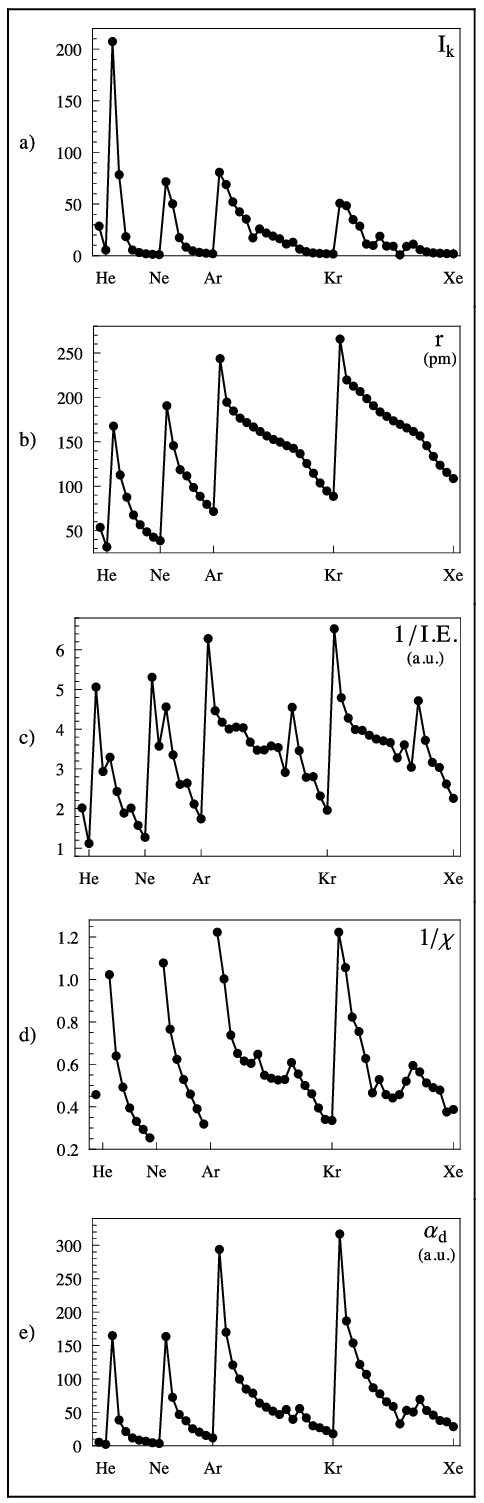}
  \caption{(a) Fisher Information in momentum space (b) Atomic Radius, (c) First Ionization Energy (inverse), (d) Electronegativity (inverse), and (e) Atomic Dipole Polarizability, as functions of the atomic number $Z=1$ to $54$.}
  \label{graph1}
\end{figure}

In Fig. \ref{graph1} we present the atomic properties, along with the Fisher information in momentum space $I_{k}$, as functions of $Z$ (for atomic number $Z=1$ to $54$).

The first atomic property is the \emph{Atomic radius}, $r$, which is a measure of the atomic size. It is usually defined as the mean or typical distance between the nucleus and the boundary of the surrounding cloud of electrons. According to quantum theory the corresponding boundary is not a well-defined physical entity, since electrons are described as probability distributions that gradually diminish as we move away from the nucleus, without a clear, sharp cutoff. Thus, there are various definitions of atomic radius, e.g. the term may apply only to isolated atoms, or also to atoms in condensed matter, covalently bound in molecules, or in ionized and excited states; and its value may be obtained through experimental measurements, or computed from theoretical models \cite{L.Pauling}. Here we use the theoretically calculated values introduced by E. Clementi, D. L. Raimondi, and W. P. Reinhardt \cite{CRR}, using SCF functions. The corresponding atomic radii values are in picometres. 

Inverse ionization energy is another property plotted in this figure. \emph{Ionization energy} (I.E.) is the minimum energy required to eject an electron out of a neutral atom or molecule in its ground state \cite{UPAC}. In general, the $n_{th}$ ionization energy is the energy required to remove the $n_{th}$ electron since the first $n-1$ electrons have been removed; the greater the ionization energy, the more difficult it is to remove an electron. Thus, it can be considered as an indicator of the \emph{reactivity} of an element. Ionization energy was formerly called \emph{ionization potential}. Here, we study the \emph{first ionization energy}, where the corresponding values of the property are given in atomic units \cite{HKK}. 

Inverse electronegativity is the third property presented. \emph{Electronegativity}, $\chi$, describes the tendency of an atom to attract electrons to itself \cite{UPAC, Pauling}. The electronegativity of an atom depends on both the atomic weight and the distance that its valence electrons reside from the charged nucleus. The higher the electronegativity number associated to an atom the more the ability of the atom to attract electrons to it. Here, we use the relative scale introduced by Pauling, the Pauling scale, where electronegativity is a dimensionless quantity \cite{Atkins}.

Finally, we plot the Atomic Dipole Polarizability. \emph{Polarizability} is a property that describes the lowest order response of the electron cloud of an atom or a molecule to an external electric field. \emph{Atomic static dipole polarizability}, $a_{d}$, is a linear response property, defined as the second derivative of the total electronic energy with respect to the external homogeneous electric field (quadratic Stark effect) \cite{UPAC}. The values of atomic static dipole polarizability used here are given in atomic units \cite{Schwerdtfeger}. 
 
It is obvious that $I_{k}$ simulates the corresponding periodicity pattern that characterizes all four atomic properties. The overall correlation is found to be excellent and implies that those properties are intimately related to momentum space. The reason that $I_{k}$ succeeds to reflect the periodic properties (of at least the first 54 values of $Z$) is because Fisher information is directly related to the local character of the probability distribution. For instance, if we are interested in momentum space, when $n(\mathbf{k})$ is more localized, the derivative $\nabla {n(\mathbf{k})}$ takes higher values and the same happens for $I_{k}$. On the other hand if $n(\mathbf{k})$ is delocalized in momentum space, then $I_{k}$ takes lower values. In the case of a noble gas, where we have closed shells atoms, the density distribution in position space $\rho(\mathbf{r})$ is more localized compared to other atoms, while $n(\mathbf{k})$ is more delocalized, especially for large values of $k$ according to Heisenberg's uncertainty principle, thus $I_{k}$ takes lower values.

\begin{figure}[!htb]
        \centering
        \begin{subfigure}[b]{0.3\textwidth}
                \centering
                \includegraphics[height=9.5cm,width=5.5cm]{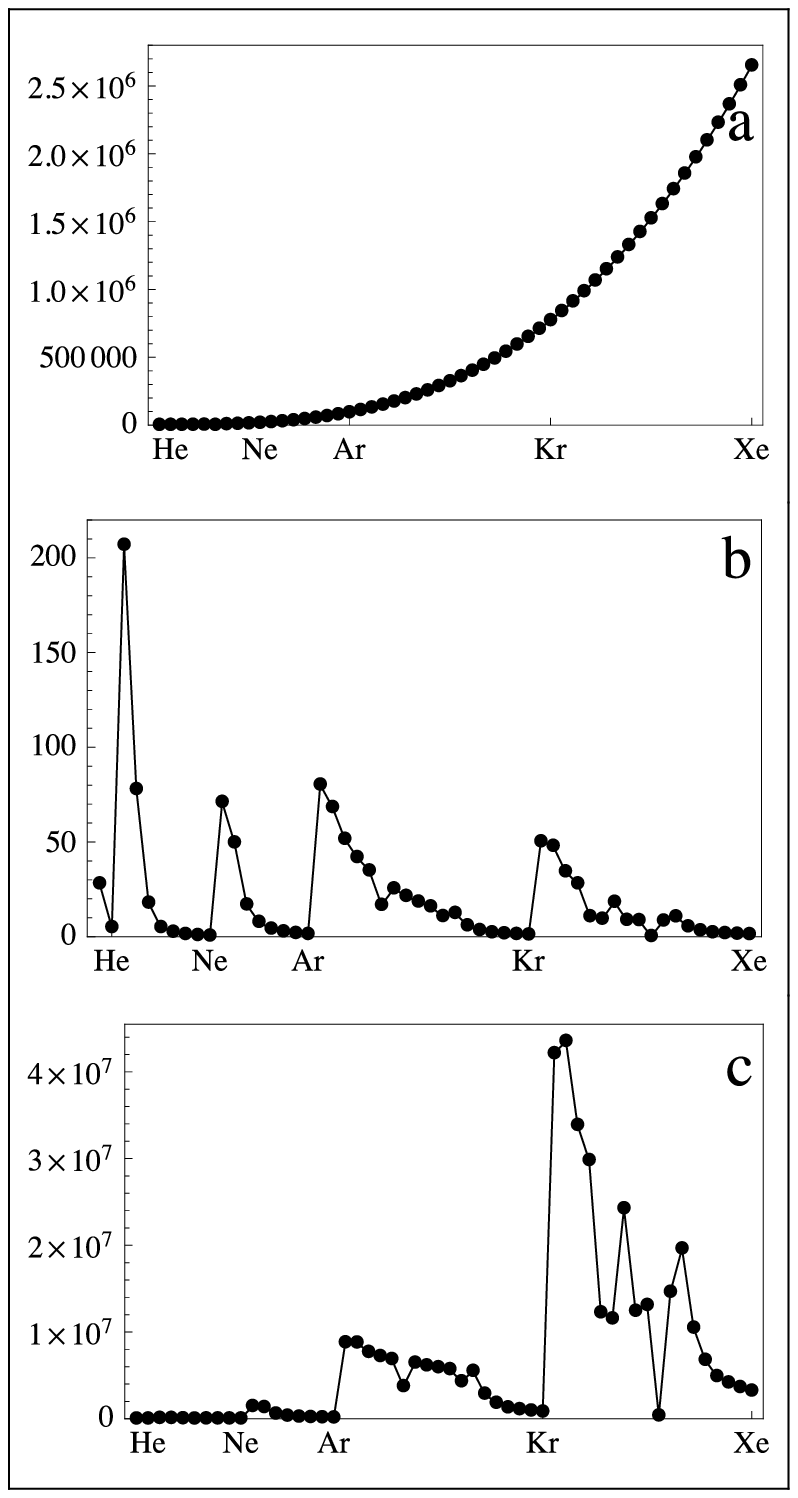}
                \caption*{Fisher Information}
        \end{subfigure}
        \begin{subfigure}[b]{0.3\textwidth}
                \centering
                \includegraphics[height=9.5cm,width=5.5cm]{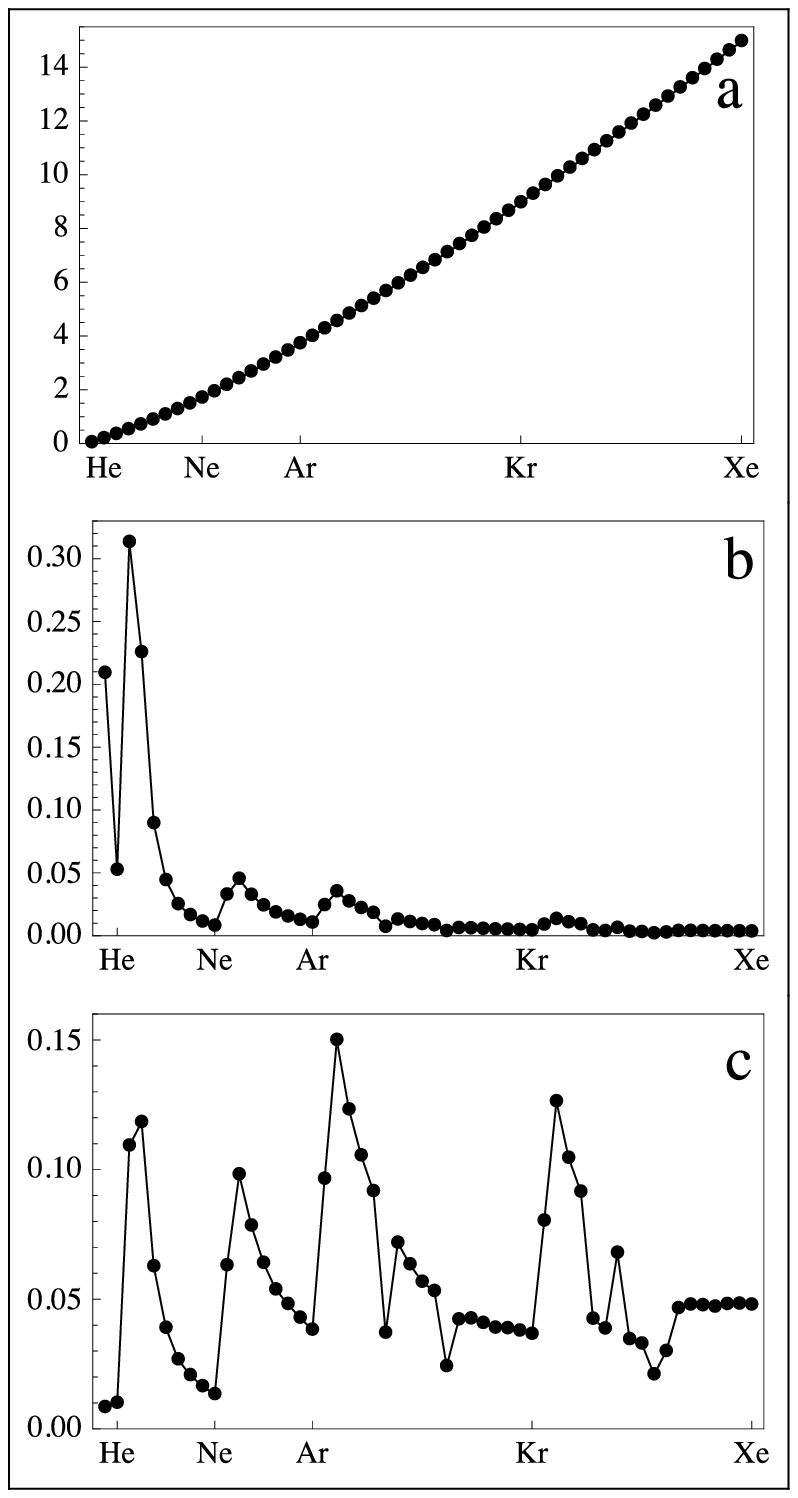}
                \caption*{Disequilibrium}
        \end{subfigure}
        \begin{subfigure}[b]{0.3\textwidth}
                \centering
                \includegraphics[height=9.5cm,width=5.5cm]{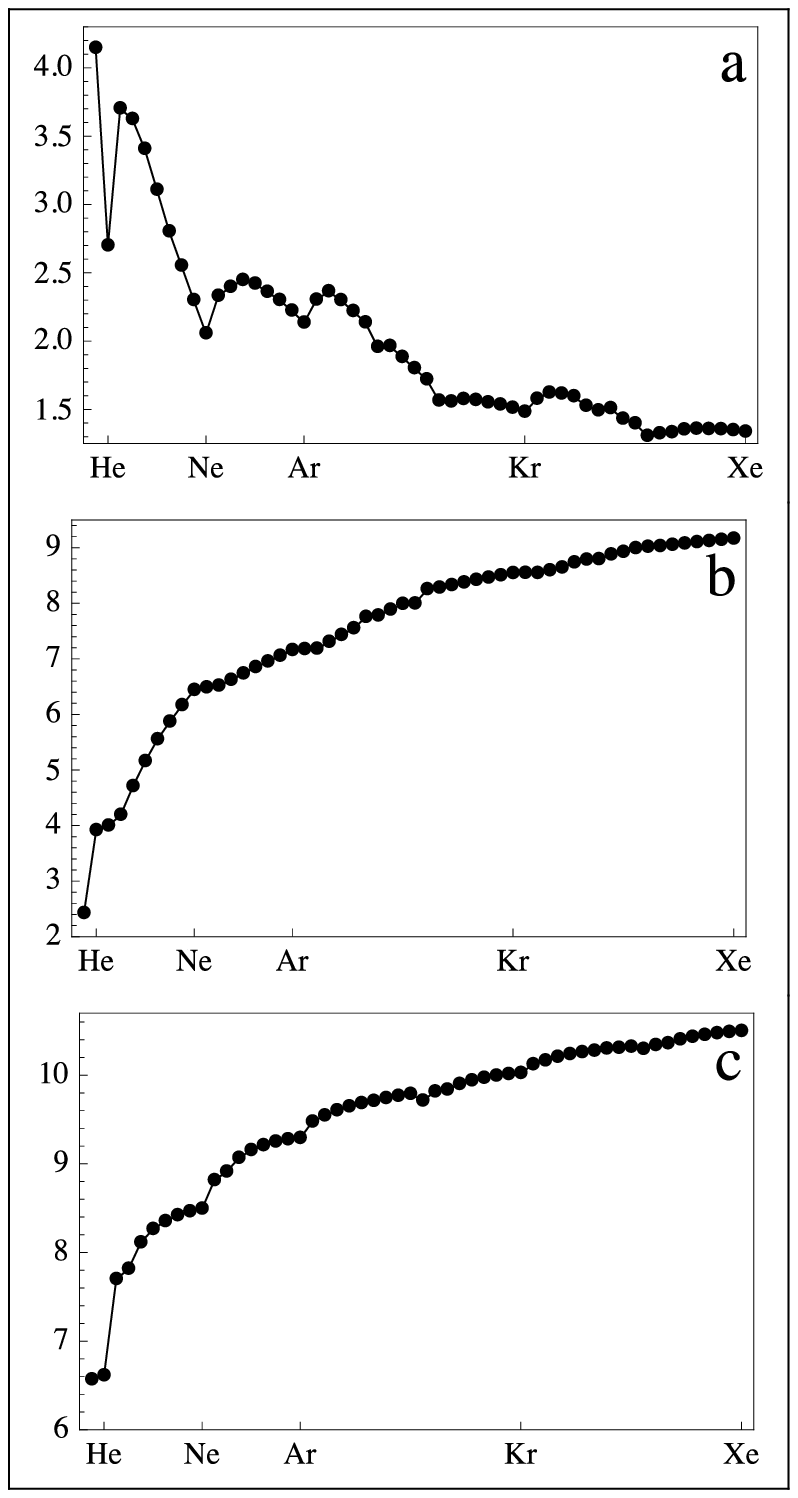}
                \caption*{Shannon Entropy}
        \end{subfigure}
\caption{Fisher Information, disequilibrium and Shannon Entropy are plotted in both (a) position and (b) momentum space, along with the corresponding (c) ``total" measure.}
\label{graph2}
\end{figure}

In Fig. \ref{graph2} we plot Fisher information, disequilibrium and Shannon information entropy, in both position and momentum spaces, along with the corresponding ``total" measure, as functions of $Z$. It is seen that besides Fisher information $I_{k}$ which, as mentioned above, provides a lot of information about the atomic structure, $I_r$ is a monotonous increasing function of $Z$, and $I_T$ just amplifies the results obtained in $I_k$ case.

For small values of $Z$ ($Z=1-18$), disequilibrium $D_{k}$ is also a sensitive index, but for elements with higher $Z$ its behavior is dominated by higher values of $Z$ and its sensitivity is lower than in the region $1\leq Z\leq 18$. In position space, similarly to $I_r$, disequilibrium $D_r$ does not seem to provide any information about the atomic structure, while the net disequilibrium $D_T$ does provide some insight about atomic structure though they appear certain deviations from the known periodicity of the elements.

The periodicity pattern that characterizes all four atomic properties is also reflected by Shannon information entropy $S_r$ in position space, but with less sensitivity compared to $I_k$. In this sense, Fisher information and Shannon information entropy present a reciprocal relation in dual spaces i.e. $I_{k}$ is complementary to $S_r$. On the other hand, $S_{k}$ and $S_T$ are almost monotonically increasing functions and seem to carry less information about atomic properties.

\begin{figure}[!htb]
        \centering
        \begin{subfigure}[b]{0.3\textwidth}
                \centering
                \includegraphics[height=9.5cm,width=5.5cm]{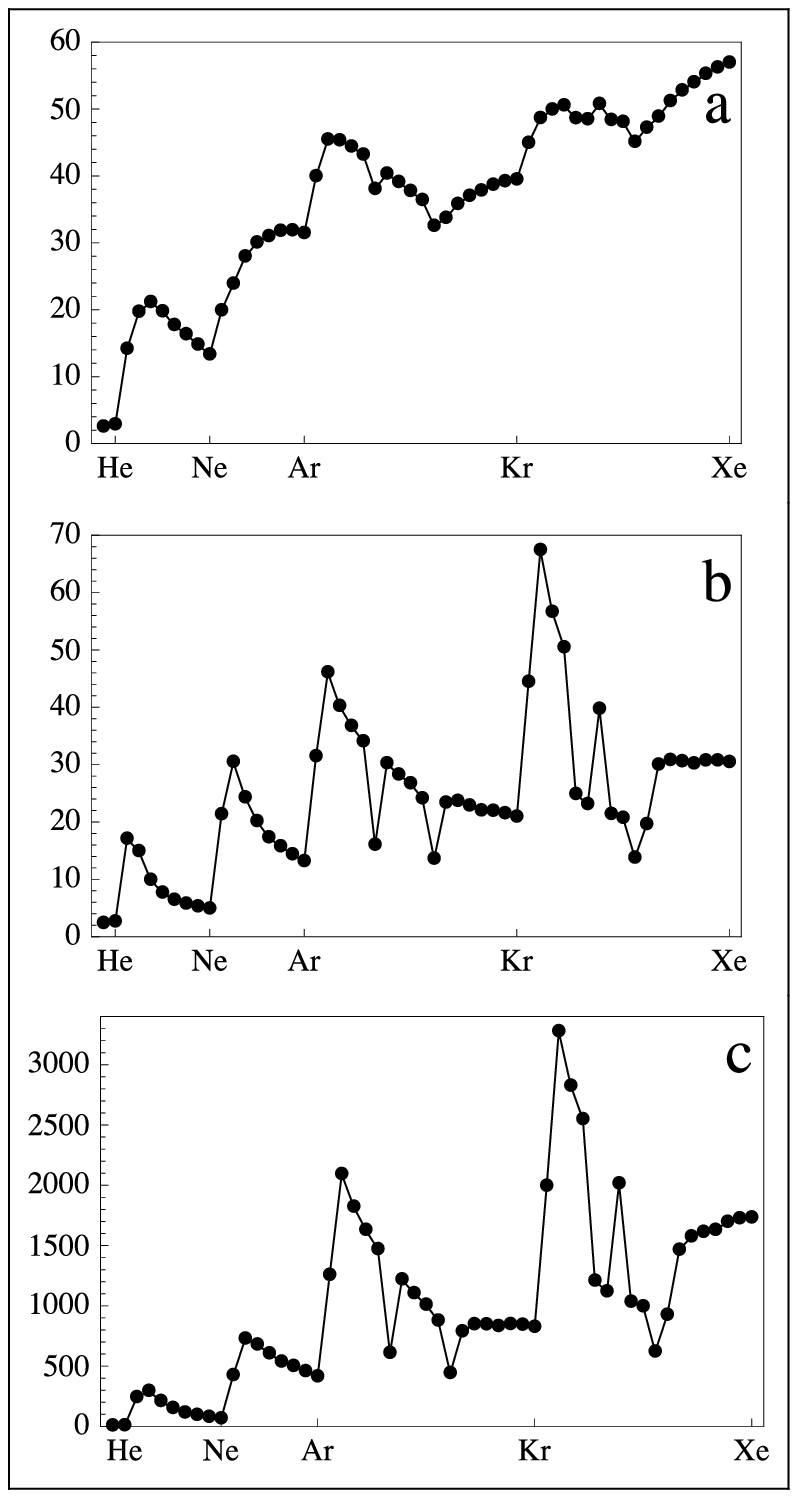}
                \caption*{LMC complexity}
        \end{subfigure}
        \begin{subfigure}[b]{0.3\textwidth}
                \centering
                \includegraphics[height=9.5cm,width=5.5cm]{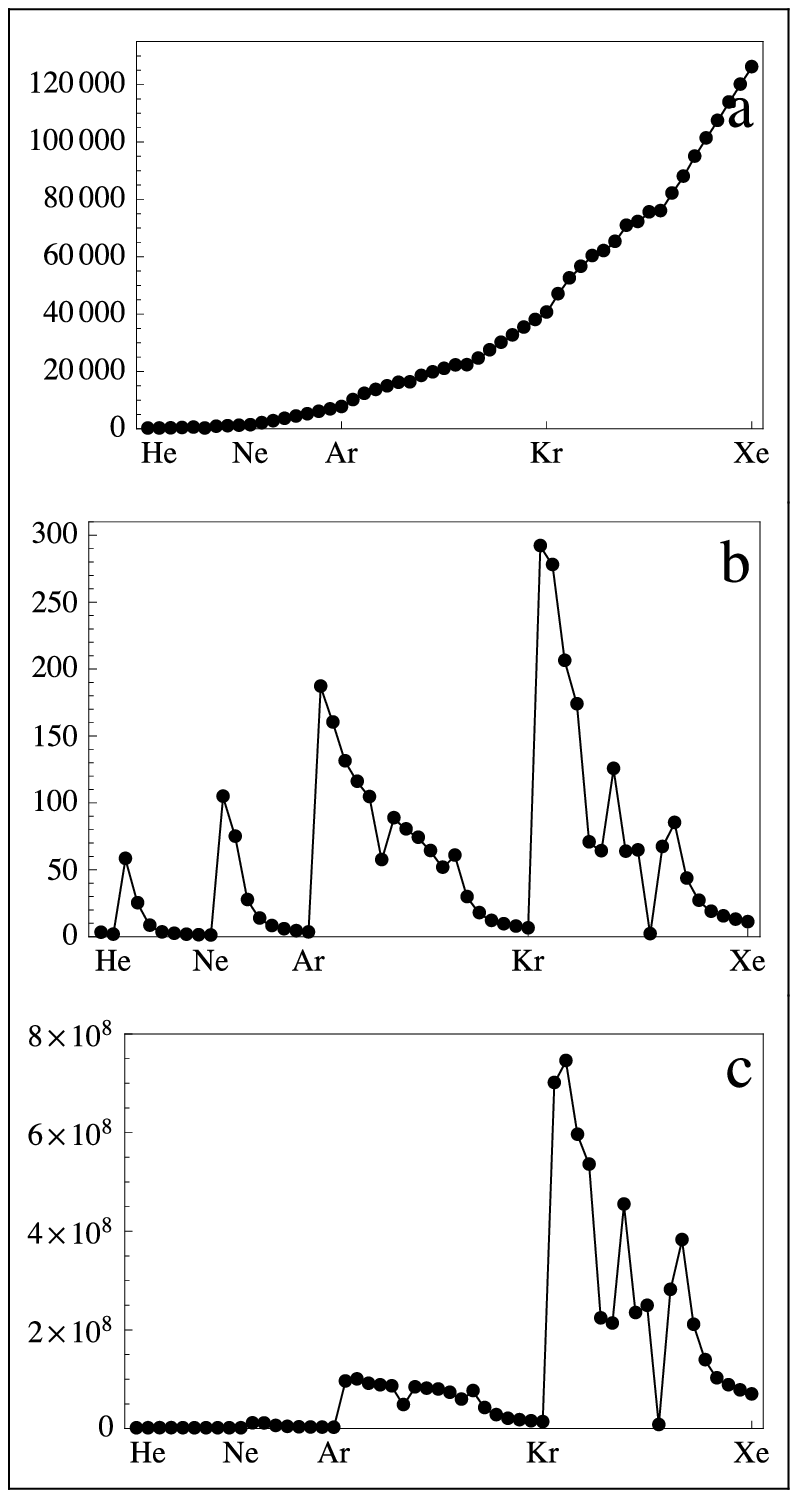}
                \caption*{Fisher-Shannon Plane}
        \end{subfigure}
\caption{LMC complexity and Fisher-Shannon plane are plotted in both (a) position and (b) momentum space, along with the corresponding (c) ``total" measure.}
\label{graph3}
\end{figure}

Finally, LMC complexity and Fisher-Shannon plane, as functions of $Z$ are plotted in Fig. \ref{graph3}. LMC complexity in momentum space reflects the aforementioned properties, something which is expected given that $C_k$ is proportional to $D_k$. In position space there is no sign of correlation, and $C_T$ shows the same pattern as $C_k$ with higher values. Fisher-Shannon plane is the last measure presented in this work. In momentum space $P_k$ reveal the atomic structure in a very detailed manner, something that is expected since $P_k$ is proportional to $I_k$. On the other hand $P_r$ is a monotonically increasing function, and $P_T$ just like $I_T$ amplifies what has already been found in momentum space.

In order to further explore possible correlations between the various information measures and the experimental quantities we employ the so-called Pearson correlation coefficient defined as

\begin{equation}
r_{XY}=\frac{\sum_{i=1}^{n} (X_i-\overline{X}) (Y_i-\overline{Y})}{\sqrt{\sum_{i=1}^{n} \left(X_i-\overline{X} \right)^2}\sqrt{\sum_{i=1}^{n} \left(Y_i-\overline{Y} \right)^2}},
\end{equation}

\noindent where, $X$ and $Y$ are two variables expected of being tied to each other, and $\overline{X}=\frac{1}{n}\sum_{i=1}^n X_i$, $\overline{Y}=\frac{1}{n} \sum_{i=1}^n Y_i$ are the mean values of $X$ and $Y$ respectively \cite{Pearson, Brandt}. The coefficient is seen as covariance of the two variables, normalized with dispersions. Actually, Pearson coefficient is well suited for picking up a linear correlation between variables. The definition imposes constraints $-1\leq r_{XY}\leq1$. The values of  $|r_{XY}|$ close to $1$ signify a very tight correlation between $X$ and $Y$, while values close to $0$ signify lack of a correlation. It is worth to point out that in our problem Pearson coefficient provides only an average correlation between the physical quantities and the relevant information measures and it is not suitable to reflect the characteristics of the system in detail.

\begin{figure}[!htb]
\centering
  \includegraphics[height=7cm,width=14cm]{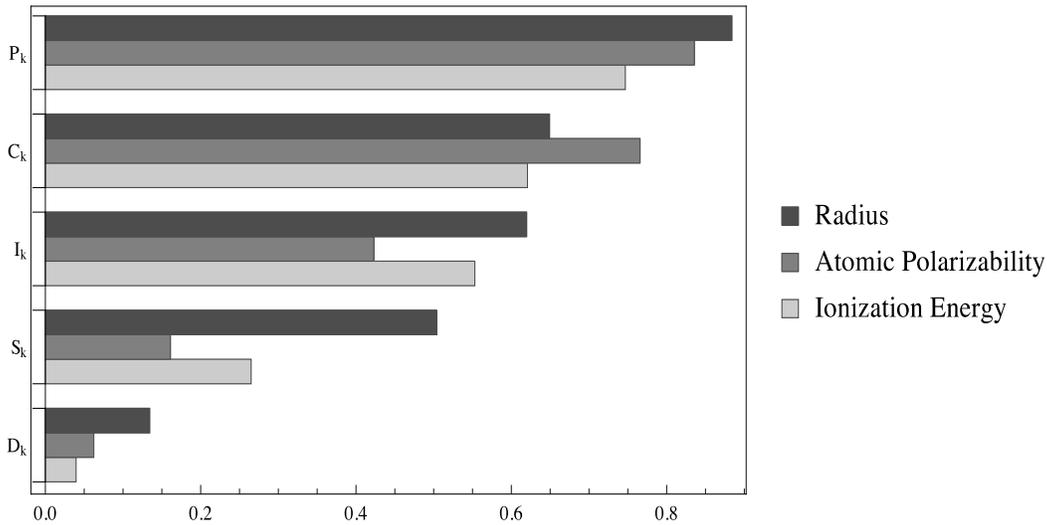}
\caption{Absolute values of Pearson coefficient for several measures and atomic properties in momentum space.}
\label{graph4}
\end{figure}

In Fig. \ref{graph4} we present a comparative bar chart of the absolute values of Pearson coefficient for several measures in momentum space and three of the aforementioned atomic properties (Radius,  Atomic Polarizability, Ionization Energy). Of course net measures also provide values very close to those presented but from Fig. \ref{graph2} it is apparent that these measures only amplify the momentum space trend. We should mention that only $D_k$ against Radius has a negative coefficient, but generally disequilibrium does not strongly correlate with any of these three experimental quantities. Shannon information also does not correlate and this is obvious also from the Pearson coefficient. On the other hand $I_k$, $C_k$ and $P_k$ have higher values, something that is also reflected in Figs. \ref{graph2} and \ref{graph3}.

The relation to momentum space stated above can be easily observed if we compare the theory of electron impact ionization with the so-called $(e,2 e)$ experiments. In such experiments, two continuum electrons emerging from the collision are detected in coincidence, and all electron momentum components measured. The observed cross-section is called triply differential and for its description a first-order approximation (called plane-wave impulse approximation) is adequate. The corresponding cross-section factorizes into the spin-averaged free electron-electron scattering cross-section $f_{ee}$ (which depends on the incoming electron momentum ${\bf k}_0$ and the outgoing electron momenta ${\bf k}_1$ and ${\bf k}_2$) and the momentum distribution of the ionized electron in its initial state (with energy eigenvalue $E_i$) $|\phi_{i}({\bf q})|^2$  and is given by the formula \cite{Keller, McCarthy}

\begin{equation}
\frac{d^3\sigma^{PWIA}}{d\Omega_1d\Omega_2 dE}=(2\pi)^4\frac{k_1k_2}{k_0}f_{ee}({\bf k}_0,{\bf k}_1,{\bf k}_2) |\phi_{i}({\bf q})|^2,
\end{equation}

\noindent where ${\bf q}={\bf k}_0-{\bf k}_1-{\bf k}_2$ is the momentum of the recoil ion. Such experiments can be considered as an ``electron momentum spectroscopy", due to the  dependence of the cross-section on the momentum distribution of the initial state. Although in principle we are able to obtain the complete quantum mechanical information on the effective-electron orbitals, in practice only the spherical average $|\phi_{i}({\bf |q|})|^2$ can be measured for atoms and molecules \cite{Keller}. 

It is clear that the ionization energy of an atom and consequently energy fluctuations that occur in the system are reflected in momentum space. More specifically, each orbital has its own density distribution so it is interesting to study the relation between Fisher Information and energy. Instead of energy we use the second moment of the density distribution in momentum space which is an energy functional and is defined as

\begin{equation}
\mu_2=\langle k^2\rangle=4 \pi \int_0^\infty k^4 n(k) \,dk.
\end{equation}

\begin{figure}[!htb]
\centering
  \includegraphics[height=7cm,width=10cm]{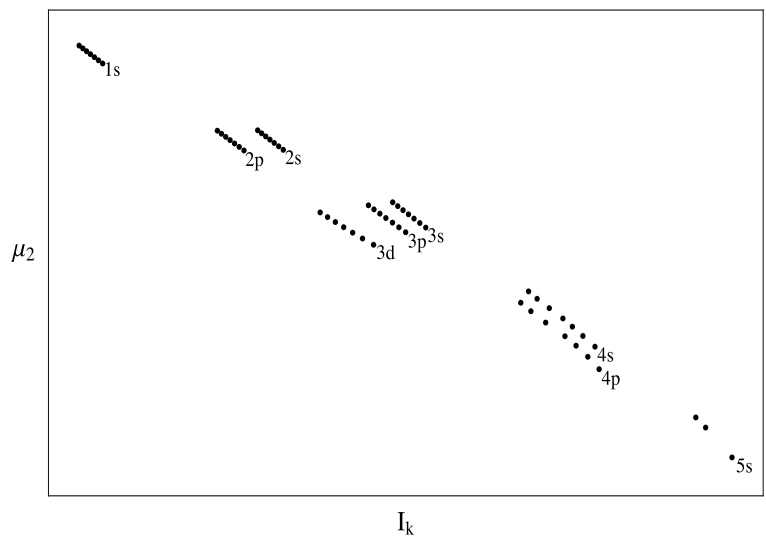}
\caption{$2nd$ moment of $n(k)$, $\mu_2$, is plotted against Fisher Information.}
\label{graph5}
\end{figure}

\noindent We plot $\mu_2$ versus $I_k$ for atoms with $33\leq Z\leq39$ in a log-log diagram (Fig. \ref{graph5}). A linear relationship in each orbital for closed shells (1s, 2s, 2p, 3s, 3p, 3d) is observed while in orbitals 4s, 4p, 5s this relation is slightly modified because there are atoms with either open shells or both closed and open shells.

\section{Summary and Conclusions}\label{Summary and Conclusions}

The main conclusion is that Fisher information exhibits a very good qualitative correlation with properties concerning the atomic structure. More specifically, $I_k$ is very sensitive to shell effects, simulating some of the most characteristic atomic properties such as the atomic radius, ionization energy, electronegativity, and atomic polarizability. That correlation is corroborated quantitatively by the Pearson coefficient.

This is a reasonable result, since Fisher information is closely related to the kinetic energy of a system, and depends on the gradient of the probability distribution, providing a local measure of smoothness, in contrast to Shannon, R\' {e}nyi or Tsalis information entropies, which provide a global way of characterizing ``uncertainty".

\onecolumngrid 
\newpage

\section{Figures}\label{Figures}

\end{document}